\documentstyle[12pt]{article}

\date{\today}

\def\be{\begin{equation}}
\def\ee{\end{equation}}
\def\bear{\begin{eqnarray}}
\def\eear{\end{eqnarray}}
\def\nn{\nonumber}

\def\bra{{\langle}}
\def\ket{{\rangle}}

\def\wdg{{\wedge}}                              

\newcommand\px[1]{{\partial_{#1}}}
\newcommand\qx[1]{{\partial^{#1}}}

\newcommand\rep[1]{{\bf {#1}}}      
\newcommand\tr[1]{{\mbox{tr}\{{#1}\}}}          
\newcommand\ev[1]{{\bra {#1} \ket}}             
\newcommand\com[2]{{\lbrack {#1},{#2}\rbrack}}  

\def\a{{\alpha}}
\def\b{{\beta}}

\def\u{{\mu}}
\def\v{{\nu}}

\def\th{{\theta}}

\def\IZ{{Z\kern-.5em  Z}}                       
\def\IR{{I\kern-.3em  R}}                       
\def\IC{{C\kern-.6em  /}}                       

\def\cO{{\cal O}} 
\def\cwO{{\widetilde{{\cal O}}}} 
\def\hO{{\hat{{\cal O}}}}
\def\wxi{{\widetilde{\xi}}}
\def\Dsl{{{D\!\!\!\!/\,\,}}} 
\def\hM{{\hat{\cal M}}} 
\newcommand\MHT[1]{{\widehat{\bf T}^{#1}}}
\def\cM{{\cal M}} 
\def\wMs{{\widetilde{M}_s}} 
\def\wG{{\widetilde{G}}} 
\def\wB{{\widetilde{B}}} 
\def\BZ{{{\bf Z}}}

\def\BR{{{\bf R}}}
\newcommand\MR[1]{{{\bf R}^{#1}}}               
\newcommand\MS[1]{{{\bf S}^{#1}}}               
\newcommand\AdS[1]{{{\bf AdS}_{#1}}}            
\newcommand\MT[1]{{{\bf T}^{#1}}}               

\newcommand\SUSY[1]{{{\cal N}= {#1}}}           

\def\lR{{R_{\|}}} 
\def\vsig{{\vec{\sigma}}} 

\begin{document}
\begin{titlepage}
\titlepage
\rightline{hep-th/0010143, PUPT-1958}
\rightline{\today}
\vskip 1cm
\centerline{{\Huge New M(atrix)-models}}
\centerline{{\Huge for Commutative and Noncommutative}}
\centerline{{\Huge Gauge Theories}}
\vskip 1cm
\centerline{
Ori J. Ganor
}
\vskip 0.5cm

\begin{center}
Department of Physics, Jadwin Hall \\
Princeton University \\
Princeton, NJ 08544, USA\\
\centerline{email: {\tt origa@viper.princeton.edu}}
\end{center}

\vskip 0.5cm

\abstract{
We propose a M(atrix) model for $\SUSY{4}$ $SU(k)$
Super-Yang-Mills theory compactified on $T^4$.
In this model it is possible to make $T^4$ noncommutative
and it is easy to turn on all 6 components of
the noncommutativity on $T^4$.
The action of S-duality on the noncommutativity
parameters is also manifest.
The M(atrix)-model is given by
the large $N$ limit of a $\sigma$-model on $T^2$
whose target space is the moduli space of $k$ $SU(N)$ instantons on
$T^3\times R$.
 We also propose that the
$SU(k)$ 2+1D $Spin(8)$ theory (the low-energy description of $k$ M2-branes)
on $T^3$ corresponds to the large $N$ limit of an integral 
over the latter instanton moduli space.
The identification is based on the fact that
Euclidean wrapped M2-branes in toroidally compactified M-theory
correspond to instantons in the M(atrix)-model.
In the new M(atrix) models, operators with nonzero
momentum along $T^3$ (or $T^4$)
correspond to insertions of Wilson lines along a 1-cycle that
is determined by the momentum.
Momentum is conserved in the large $N$ limit.
}
\end{titlepage}
             

\section{Introduction}\label{intro}
The M(atrix)-model for M-theory \cite{BFSS} uses the large $N$ 
limit of 0+1D  $U(N)$ Super-Yang-Mills theory (SYM) with $\SUSY{16}$
supersymmetry (see \cite{CH,WHN,WitDBR} for
previous appearances of this theory).
M(atrix)-theory also suggests a new nonperturbative formulation 
of various field theories. Thus, the $(2,0)$-theory that is realized
as the low-energy limit of $k$ coincident M5-branes is described
in M(atrix)-theory by the large $N$ limit of the $\SUSY{8}$
supersymmetric extension of quantum mechanics on 
the moduli space $\cM_{N,k}$ of $N$ $U(k)$ instantons on
$\MR{4}$ \cite{ABKSS,WitQHB,ABS}. The M(atrix)-model of 3+1D $\SUSY{4}$
SYM and the 2+1D $Spin(8)$ theory \cite{SetSus,BanSei} can be derived
from a limit of the moduli space of $N$ $U(k)$ instantons on 
$\MT{2}\times\MR{2}$ and $\MT{3}\times\BR$ \cite{GanSet}.

In all of those M(atrix)-models the field theory on $\MR{d,1}$ is required
to have at least two noncompact directions, one of which is light-like.
Alternatively, the theory can have a compactified light-like direction
as in \cite{Sussk} but it cannot have all $d$ space-like directions
compactified simultaneously.
Another question that can only partially be answered in the framework
of those M(atrix)-models is how to describe the field theories
on a noncommutative space.

Standard gauge field-theories have an extension to noncommutative
spaces for which the product of fields is replaced with a noncommutative
$\star$-product. These theories are parameterized by an anti-symmetric 
contravariant 2-tensor (bivector) $\th^{ij}$.
As will be explained in section (\ref{noncom}),
using the ideas of \cite{ABS,NekSch,Berk},
the M(atrix)-models of \cite{GanSet} can easily be extended 
to describe $\SUSY{4}$ SYM on a noncommutative $\MR{3,1}$ (NCSYM).
However, only two out of the 6 components of $\th^{ij}$
can easily be turned on in this framework.

In these notes we would like to suggest M(atrix)-models for
completely compactified, Euclidean field-theories.
The theories that we will discuss are $\SUSY{4}$ SYM compactified 
on $\MT{4}$ and the $Spin(8)$ theory compactified on $\MT{3}$.
The M(atrix)-models are going to be a $\sigma$-model on $\MT{2}$
with  a certain moduli space of instantons as the target space,
and an integral on a certain moduli space of instantons, respectively.
In these models for $\SUSY{4}$ SYM all 6 components of the 
noncommutativity can easily be turned on.

The idea behind these models is as follows.
Starting with M-theory compactified on $\MT{d}$
we can consider  Euclidean M2-branes or M5-branes 
that wrap 3-cycles or 6-cycles of $\MT{d}$ and produce instanton
effects (see \cite{OPRev} and refs therein).
In general, such instanton effects can be separated from other
quantum effects because they are accompanied by a characteristic
phase dependence on the moduli of the compactification.
For example, if we compactify M-theory on $\MT{3}$, the 3-form
flux along $\MT{3}$, $\phi\equiv\int_\MT{3} C$, is a periodic modulus.
An instanton effect that is a result of $k$ Euclidean M2-branes
wrapping $\MT{3}$ comes with the characteristic prefactor
$e^{i k\phi}$. In addition to the phase, the instanton effect
can be calculated from the action of the branes and the specific
physical question (e.g. the scattering amplitude of 
some particles) that we are trying to solve.
For example, when all the dimensions of $\MT{3}$ are large compared
to $M_p^{-1}$, the instanton contribution can presumably be calculated
from the low-energy theory of $k$ coincident M2-branes, i.e.
the 3D $Spin(8)$ theory \cite{SetSus,BanSei}.

How is all of that manifested in M(atrix)-theory?
The M(atrix)-model of M-theory on $\MT{3}$ is 3+1D $\SUSY{4}$
SYM compactified on $\MT{3}$.
The instanton described above corresponds to an instanton
of the SYM theory \cite{GRT} and therefore there must be
a map from quantities calculated in the $Spin(8)$ theory to
quantities calculated by integrating over the moduli space of
instantons.

Thus, the large $N$ limit of the integral over the moduli space
of instantons on $\MT{3}\times\BR$, where $\BR$ is the time
direction, is a M(atrix)-model for the compactified $Spin(8)$ theory.
In a similar spirit, one can 
derive a M(atrix)-model for the compactified $\SUSY{4}$ SYM.
The purpose of these notes is to derive these M(atrix)-models
and explore them.

The paper is organized as follows.
In section (\ref{models}) we construct the new M(atrix)-models
for the $Spin(8)$ theory on $\MT{3}$ and for $\SUSY{4}$ SYM on $\MT{4}$
and argue for the decoupling of other
M(atrix)-theory fields from the variables in the moduli spaces
of instantons. 
In section (\ref{flat}) we discuss the flat directions
of the field theories and the instanton moduli spaces.
In that section we also describe compactification of the field
theories with R-symmetry twists.
 Their M(atrix)-models are related to dipole-theories \cite{CGK,BG}
(and see also \cite{BanMot} for a special case).
In section (\ref{maps}) we attempt to construct a map between 
questions about the compactified theory and questions about the
new M(atrix)-models.
We propose that field theory momentum should be translated into
$\BZ_N$ charge in the M(atrix) model and operators that carry momentum
should be mapped to insertions of Wilson lines.
Finally, in section (\ref{noncom}) we describe the extension of the
$\SUSY{4}$ SYM M(atrix)-model to a noncommutative $\MT{4}$.

For the benefit of readers who wish to skip
sections (\ref{flat})-(\ref{maps}),
we note that section (\ref{noncom}) can be read immediately after
section (\ref{models}).


\section{New M(atrix)-models from instantons}\label{models}
We will derive the M(atrix)-models for the 2+1D $Spin(8)$
theory on $\MT{3}$ and the 3+1D $\SUSY{4}$ SYM on $\MT{4}$
by studying instantons in M-theory on $\MT{3}$ or type-IIB string
theory on $\MT{4}$. We will assume that the metric is Euclidean
in both cases.

\subsection{The 3D $Spin(8)$ theory on $\MT{3}$}
M-theory on $\MT{3}$ has a moduli space:
$$
(SL(3,\BZ)\backslash SL(3,\BR)/SO(3))\times
(SL(2,\BZ)\backslash SL(2,\BR)/SO(2)).
$$
The first factor corresponds to the geometrical shape of the $\MT{3}$
and the second factor parameterizes the volume and 3-form flux
such that the combination:
$$
\tau = \frac{i M_p^3 V + C}{2\pi},
$$
(where $V$ is the volume, $M_p$
is the 11D Planck energy scale and $C$ is the 3-form flux)
transforms under the generators of
$SL(2,\BZ)$ as $\tau\rightarrow\tau+1$ and
$\tau\rightarrow -1/\tau$.
The M(atrix)-model of M-theory on $\MT{3}$ is \cite{BFSS,Wati,GRT}
$\SUSY{4}$ $U(N)$ SYM on the geometrical dual $\MHT{3}$ with
coupling constant and $\theta$-angle given by the same 
$\tau = \frac{4\pi i}{g^2} + \frac{\theta}{2\pi}$.
The volume of $\MHT{3}$ is immaterial because $\SUSY{4}$ SYM
is a conformally invariant theory.

Various amplitudes in M-theory on $\MT{3}$ receive
instanton contributions from Euclidean M2-branes wrapped on
$\MT{3}$ (see \cite{OPRev} and refs therein).
The amplitude of an instanton effect made out of $k$ M2-branes
has the characteristic factor $e^{i k C}$.
In the M(atrix)-model, such phases correspond to effects
of $k$ Yang-Mills instantons.

Now consider the limit $M_p^3 V\rightarrow \infty$.
In this limit, when the $k$ M2-branes are close to each other,
the instanton effect has a factor $e^{2\pi i k\tau}Z$ where
$Z$ can be calculated from the partition function of the
``$U(k)$'' $Spin(8)$ theory \cite{SetSus,BanSei}.
In the M(atrix)-model, this limit corresponds to 
$g_{YM}\rightarrow 0$ and the effect we are looking for 
comes from a sector with Yang-Mills instanton number $k$
that carries a prefactor $e^{2\pi i k\tau}$.
If we also take the low-energy limit on both sides,
then in M-theory gravity decouples and we get the $Spin(8)$ theory.
In M(atrix)-theory the instanton contribution is of the form
$e^{2\pi i k\tau}Z'$ where $Z'$ can be calculated
from an integral over the moduli space of $k$ $SU(N)$ instantons
on $\MHT{3}\times\BR$. Here $\BR$ is the time direction.

So, the M(atrix)-model for the ``$SU(k)$'' $Spin(8)$ theory compactified
on $\MT{3}$ is an integral over $\cM_{N,k}(\MHT{3}\times\BR)$ --
the moduli space of $k$ $SU(N)$ instantons on the dual $\MHT{3}$
times $\BR$ in the limit $N\rightarrow\infty$.

\subsection{4D $\SUSY{4}$ SYM on $\MT{4}$}\label{mmsym}
We can use a similar reasoning to find a M(atrix)-model
for $\SUSY{4}$ $SU(k)$ SYM on $\MT{4}$.
We start with type-IIB string theory on $\MT{4}$ which, for
simplicity, we take to be rectangular with radii $R_1,\dots,R_4$.
We let the complex type-IIB coupling constant be $\tau$.
This is equivalent to M-theory on $\MT{3}\times\MT{2}$ where
$\MT{2}$ has complex structure $\tau$ and area 
$\tau_2^{-1/3}M_p^{-2} (M_s R_4)^{-4/3}$
($M_p$ is the 11D Planck scale and $M_s$ is the 10D string scale).
$\MT{3}$ has radii 
$M_p^{-1} M_s^{4/3}\tau_2^{1/3}R_4^{1/3} R_i$ ($i=1\dots 3$).
According to \cite{SeiVBR}, the M(atrix)-model for M-theory
on $\MT{5}$ with these parameters is given by the 
(type-IIB) little-string-theory
(LST) compactified on another $\MHT{3}\times\MT{2}$.
We define  the LST energy scale to be $\wMs$.
The $\MT{2}$ has  area:
\be\label{ATtwo}
A_2\equiv \wMs^{-2} M_s^4\tau_2 R_1 R_2 R_3 R_4,
\ee
and $\MHT{3}$ has radii:
\be\label{LiRad}
L_i \equiv \wMs^{-1} R_4^{-1/2} (R_1 R_2 R_3)^{1/2} R_i^{-1}
\qquad i=1\dots 3,
\ee
so that the area of $\MHT{3}$ is:
\be\label{ATthree}
A_3\equiv \wMs^{-3} R_4^{-3/2} (R_1 R_2 R_3)^{1/2}.
\ee
Now we take the limit $R_1,R_2,R_3,R_4\gg M_s^{-1}$.
In this limit:
$$
\wMs^3 A_3 = O(1),\qquad \wMs^2 A_2  \gg 1.
$$
The periodic modulus that is related to wrapped D3-branes
in type-IIB, i.e. the integral of the 4-form RR-field
over $\MT{4}$, becomes the integral of the 3-form M-theory
field over $\MT{3}$. Therefore, in the LST $k$ becomes
the instanton number on $\MHT{3}\times\BR$, where $\BR$ is the time
direction.

We wish to extract from this LST the part that describes
the M(atrix)-model for 4D $\SUSY{4}$ SYM on $\MT{4}$.
As we have seen, the LST is formulated on 
$W\equiv\MT{2}\times\MHT{3}\times\BR$
where $\MHT{3}$ is of order $\wMs^{-1}$ and $\MT{2}$ is large.
We would like to argue that this M(atrix)-model is the
partition function of a certain 2D CFT on $\MT{2}$ with
complex structure $\tau$. This $\MT{2}$ corresponds to the first
factor in $W$. The CFT, we will argue, is a $\sigma$-model
with target space $X\equiv\cM_{N,k}(\MHT{3}\times\BR)$ --
the moduli space of $k$ $SU(N)$ instantons on $\MHT{3}\times\BR$.

To describe the $\sigma$-model we need to describe a metric on $X$
and a $B$-field that takes values in $H^2(X,\BR)/H^2(X,\BZ)$.
Let us first describe the parameters of $\MHT{3}$ more generally.
Starting with 4D $\SUSY{4}$ SYM on $\MT{4}$, the geometrical shape
of $\MT{4}$ corresponds to a point in the moduli-space 
$SL(4,\BZ)\backslash SL(4,\BR)/SO(4)$.
This is equivalent to 
$SO(3,3,\BZ)\backslash SO(3,3,\BR)/(SO(3)\times SO(3))$.
A point in the latter space can be interpreted as a metric and
anti-symmetric 2-form B-field on $\MHT{3}$.
Thus, $\MHT{3}$ comes naturally with a metric and B-field on it.
{}From this metric and B-field we can derive a metric and B-field on
$X$ as follows. Let $\xi\in X$ denote a particular point in the 
instanton moduli space. Let $A(\xi)$ be an $N\times N$ matrix-valued
1-form on $\MHT{3}\times\BR$ that describes the instanton solution
up to a gauge transformation. Let $d_X = 4 (k-1)(N-1)$
be the dimension of $X$ (we assume that the gauge fields are zero
at $\pm\infty$).
The metric on $X$ can be written as:
\be\label{wGonX}
\wG_{ij} d\xi^i d\xi^j = \sum_{1\le i,j\le d_X}
\left(\int_{\MHT{3}\times\BR}g^{\a\b}\tr{\frac{\delta A_\a}{\delta\xi^i}
  \frac{\delta A_\b}{\delta\xi^j}}\right) d\xi^i d\xi^j,
\ee
where $g^{\a\b}$ is the metric on $\MHT{3}\times\BR$ and one
has to gauge fix $A(\xi)$ such that:
$$
D^\a \frac{\delta A_\a}{\delta\xi^i} = 0,\qquad
D^\a \equiv \qx{\a} - i\com{A^\a}{\cdot}.
$$
The 2-form on $X$ can be written as:
\be\label{wBonX}
\wB_{ij} d\xi^i \wdg d\xi^j = \sum_{1\le i,j\le d_X}
\left(\int_{\MHT{3}\times\BR}B\wdg\tr{\frac{\delta A}{\delta\xi^i}
  \wdg\frac{\delta A}{\delta\xi^j}}\right) d\xi^i \wdg d\xi^j,
\ee
The idea behind these expressions is as follows
(see also \cite{SeiVBR,ABKSS,WitQHB,GanSet}).
At low-energies and when the size of $\MHT{3}$ is large compared
to $\wMs^{-1}$ the LST can be approximated by 5+1D SYM on
$\MT{2}\times\MHT{3}\times\BR$. This action contains
$$
\int_{\MT{2}\times\MHT{3}\times\BR}\left\lbrack
\frac{M_s^2}{4}\tr{F^2} + B\wdg\tr{F\wdg F}\right\rbrack.
$$
When the size of $\MT{2}$ is very large, this action can be reduced
to the $\sigma$-model on $X$ with metric $\wG$ and 2-form $\wB$.
We conjecture that formulas (\ref{wGonX}-\ref{wBonX}) continue
to hold even when $\MHT{3}$ is small.
Note also that if the parameters $\xi_i$ are chosen to have periodicities
of order $1$ then $\wG$ and $\wB$ are independent of $\wMs$, as they should.

\subsection{Remarks}
Let us comment on the treatment of singularities in $\cM_{N,k}$.
In general, these moduli spaces of instantons have singularities at 
(real) codimension-4 that, roughly speaking, correspond to instantons
that collide. 
A particular treatment of these singularities was suggested in \cite{ABS}
in the context of the M(atrix)-model for the $(2,0)$ and LST.
There it was suggested to deform the moduli space to a smooth space by
a parameter $\theta$. It was argued that this describes a deformation
of the $(2,0)$ theory that breaks Lorentz invariance.
This deformation of $\cM_{N,k}$ was then interpreted in \cite{NekSch,Berk}
as the moduli space of  instantons on noncommutative $\MR{4}$.
The Lorentz-invariant $(2,0)$-theory or LST can be recovered in the limit
$\theta\rightarrow 0$.
In our case, we can use a similar idea to treat the singularities.
We can turn on a noncommutativity on $\MHT{3}$, which also
forces us to work with a $U(N)$ gauge group.
(For another example of how some singularities are smoothed for
noncommutative instantons on $\MT{4}$, see \cite{GaMiSa}.)

According to \cite{CDS}, the M(atrix)-model of M-theory on $\MT{5}$ with 
a nonzero 3-form flux in the light-like direction and two of the $\MT{5}$
directions is given formally by 5+1D SYM on a noncommutative $\MT{5}$
(interpreted as LST as in \cite{SeiVBR}).
The point is that in the large $N$ limit, the effect of the C-field flux
should disappear, since it can be gauged away.
Thus, in our case, changing $\MHT{3}$ to a noncommutative space
should not affect the large $N$ limit.
In our case, unlike the case of \cite{ABKSS}, smoothing out the
singularities in $\cM_{N,k}$ by turning on noncommutativity
does not result in a breakdown of Lorentz invariance and the large $N$
limit should, presumably, be unaffected!

Finally, note that the $Spin(8)$ R-symmetry of the 3D theory
and the $Spin(6)$ R-symmetry of $\SUSY{4}$ SYM are not manifest in
the new M(atrix)-models. Presumably, they are restored in the large
$N$ limit like Lorentz invariance in M(atrix)-theory \cite{BFSS}.

\subsection{Decoupling arguments}\label{xfields}
Let us consider M-theory on $\MT{3}$ again in the limit that
the dimensions of $\MT{3}$ are much bigger than $M_p^{-1}$. 
As we have seen, the sector with $k$ instantonic M2-branes
wrapping $\MT{3}$ is described in M(atrix)-theory by the sector
with instanton number $k$ in $\SUSY{4}$ SYM.
We have therefore argued that the $Spin(8)$ theory that describes
$k$ M2-branes at low-energy corresponds to the large $N$ limit
of an integral over the moduli space on instantons $\cM_{N,k}$
on $\MT{3}\times\BR$.
Let us elaborate on why we can restrict to the moduli space
of instantons and ignore the other modes of the $\SUSY{4}$ SYM
M(atrix)-model.

Let us first recall how the decoupling
occurs in the M(atrix)-models of 
\cite{ABKSS,WitQHB}.
These models describe the $(2,0)$-theories on
$\MR{6}$ as the large $N$ limit of quantum-mechanics on
a certain moduli space of instantons.
When the M(atrix)-models are derived from the M(atrix)-model
of M-theory with $k$ M5-branes \cite{BerDou},
the moduli space of instantons is obtained as the minimum of
the potential.
The excitations that take the variables
out of the moduli space of instantons are very massive.

In our case, we need a related but somewhat different argument.
We are considering the limit that $M_p^3 V\rightarrow \infty$, where
$V$ is the volume of $\MT{3}$. This corresponds to $g_{YM}\rightarrow 0$.
Let us consider a particular $U(N)$ instanton configuration.
The M(atrix)-model modes that we have neglected are the 
6 adjoint scalars $X^I$,
4 adjoint Dirac fermions on $\MT{3}\times\BR$ and the fluctuations
of the gauge fields transverse to the moduli space.
In the limit $g_{YM}\rightarrow 0$ we can keep only the quadratic
interactions and ignore the cubic and quartic terms in the Lagrangian
of $\SUSY{4}$ SYM.
Thus, we only need to show that the determinants coming from integrating
the quadratic modes in the fields of the instantons cancel between
bosons and fermions and this easily follows from supersymmetry.

\section{Flat directions and twists}\label{flat}
In section (\ref{models}) we suggested various
correspondences between partition functions of 
field-theories and integrals or $\sigma$-models
over moduli-spaces of instantons.
However, the correspondence is incomplete because 
the partition functions of the field theories are ill-defined.
The integrals defining the partition functions
have noncompact bosonic zero modes.
For the $Spin(8)$ theory there are $8k$ such modes corresponding
to the separation of the M2-branes.
For $\SUSY{4}$ SYM there are $6k$ such modes.
Moreover, there are fermionic zero modes as well.
Our task in this section is to understand how to treat those zero
modes in the corresponding M(atrix)-models.

\subsection{Regularizing the partition function}\label{regpf}
It is not difficult to trace the zero modes of the field-theory
into the M(atrix)-model.
For concreteness, let us concentrate on the $Spin(8)$ theory.
Let us start with a single Euclidean  M2-brane in
M-theory on $\MT{3}$ as in section (\ref{models}).
It has 8 bosonic zero modes corresponding to translations in
transverse directions. Two out of the eight correspond to
translations in the light-like $x^{+}$ and $x^{-}$ directions.
In M(atrix)-theory, the $x^{+}$ coordinate is periodic with
period $2\pi \lR$ and so this zero mode gives an overall finite
$2\pi \lR$ factor to the partition function of the M2-brane.
The $x^{-}$ direction corresponds  in M(atrix)-theory to
the time direction. 
The M(atrix)-model configuration was given by a $U(N)$ instanton 
on $\MHT{3}\times\BR$.
So the $x^{-}$ zero mode corresponds to translation of the position
of the instanton in the $\BR$, i.e. time, direction.
The other 6 zero modes correspond to the 6 scalars of $\SUSY{4}$ SYM.

Now consider two Euclidean M2-branes in M-theory on $\MT{3}$.
There are the overall center of mass flat directions that can be treated as
before but there are also flat directions corresponding to the relative
separation of the instantons.
The 6 flat directions that describe separation of the M2-branes
in transverse directions (orthogonal to $x^{+}$ and $x^{-}$)
correspond to breaking the $U(N)$ gauge-group by VEVs of the
6 adjoint scalars of SYM down to $U(N_1)\times U(N_2)$ 
such that $N=N_1+N_2$ and then embedding one instanton in $U(N_1)$
and the other in $U(N_2)$.
Turning on the VEVs of the scalars is
undesirable for us, since we wish to argue that we can 
restrict to the SYM instanton moduli space variables.
In order to suppress the excitations of the 6 adjoint scalars,
we have to put the instantons in some physical context.

We can imagine that we are trying to calculate the instanton
contribution to the scattering of gravitons or to some terms
in the low-energy effective action. For example, as shown in
\cite{PioKir} (see also \cite{GG}),
there is an $R^4$ correction to the low-energy
Wilsonian effective action of M-theory on $\MT{3}$ that receives
contributions from such instantons.
In a scattering amplitude that is calculated from this Wilsonian
effective action there can be two types of contributions from such
terms at instanton number 2.
One is an ``irreducible'' term where there is a single $R^4$ vertex
and we take the instanton number 2 contribution to that vertex.
The second type is a ``reducible'' contribution where there are, say,
two $R^4$ vertices connected by graviton propagators and each
vertex has an instanton number 1 contribution.
It is clear that the terms we would like to consider are those of
the irreducible type. On the field-theory side, they are calculated
from the partition function of the $SU(2)$ $Spin(8)$ theory.
the overall center of mass flat direction decouples in that partition 
function. The flat direction that corresponds to separation of 
the two M2-branes seems to give an infinite contribution,
but there are also fermionic zero modes that, if treated properly,
cancel the infinite contribution from the bosonic flat directions.
An explicit calculation was performed in \cite{GG} for the 
case of the D(-1)-instanton. 
The explicit calculation from the $R^4$ terms in type-IIB string theory
was shown to agree with the result of $\frac{5}{4}$ calculated 
from the $SU(2)$ SYM partition function in \cite{Yi,SetSte}.
It is tempting to identify this ``irreducible'' contribution on
the M-theory (compactified on $\MT{3}$) side
with the integral over the moduli space of instantons on 
$\MT{3}\times\BR$ on the M(atrix)-model side, where we set the 6 adjoint
scalar fields to zero.
On the M-theory side, the $Spin(8)$ theory is the correct description
of the dynamics of the M2-branes when the separation between the M2-branes
is much smaller than the 11D Planck length $M_p^{-1}$.
On the M(atrix)-model side this means that we should indeed
set the adjoint scalar fields to zero. (In principle we might want
to neglect only the quartic term but keep the quadratic term in the
6 scalar fields, but the contribution of the fermions
cancels the quadratic integral of the bosons.)

Ideally, we would be led to
believe that the partition function of the $Spin(8)$ theory
(which is finite once the fermionic zero modes are properly treated
to cancel the infinite integration range over the flat directions)
is equal to the integral over the moduli space of instantons
on $\MT{3}\times\BR$.
The latter integral contains an unbound flat direction that corresponds
to large separation of the instantons in the $\BR$ direction.
On the M-theory side this corresponds to separation of the M2-branes
in the $x^{-}$ direction. Once the zero modes
are treated correctly the integral over the instanton moduli space
is finite. 
In fact, the $Spin(8)$-theory partition function and the moduli space
integral need not be equal, but their ratio will determine the overall
normalization.
We will return to this point in subsection (\ref{pf}).

\subsection{Twisting}\label{subtwist}
Another way to get rid of the $8k$ flat directions is
to use twisted boundary conditions.
This is a well-known way to avoid zero-modes
(see \cite{WitLG}). In the context of M-theory, the compactification
described below was used
in \cite{GanXi} to propose a definition for a partition
function of gravity.
It will be further explored in \cite{toApp}.
The idea is to modify $\MT{3}\times\MR{8}$ (for the moment,
we  either think of space-time as Euclidean or we set one of the
$\MT{3}$ directions to be time-like) into an $\MR{8}$-fibration
over $\MT{3}$ such that when we go around a cycle of $\MT{3}$ we
perform a $Spin(8)$ rotation of the transverse $\MR{8}$.
Locally this space is flat but globally it differs from
$\MT{3}\times\MR{8}$ because of the $Spin(8)$ twists.
For generic twists, an M2-brane that wraps the $\MT{3}$ has minimal volume
when it is at the origin of each of the $\MR{8}$-fibers.
If it wishes to escape the origin it must increase its volume
so the flat directions are gone. The fermionic zero modes also become
massive, in general.
In the M(atrix)-model setting we can only use $Spin(6)$ twists
since we wish to preserve the light-like directions 
$x^{+}$ and $x^{-}$.\footnote{Two possible ways to use the whole $Spin(8)$
twists will be explored in \cite{toApp}. One is to include $SO(1,1)$
twists which is a symmetry only at the limit $N=\infty$. Another
way is to use the IKKT model\cite{IKKT}.}
The $Spin(6)$ elements must all commute with each other and
we can choose them in a subgroup $U(1)^3\subset Spin(6)$.
We will denote the $\MT{3}$ directions by $x_1,x_2,x_3$.
Let the corresponding M(atrix)-theory fields be $X_1,X_2,X_3$.
They are adjoint scalars.
$x_4,\dots,x_9$ will denote the 6 transverse directions.
Let 
$$
z_1=x_4+i x_5,\,
z_2=x_6+i x_7,\,
z_3=x_8+i x_9,
$$
and let the corresponding M(atrix) theory fields be $Z_1,Z_2,Z_3$.
Let the twists be given by the identification:
\bear
(x_1,x_2,x_3,z_1,z_2,z_3) &\sim&
(x_1 + 2\pi n_1 R_1,\,
 x_2 + 2\pi n_2 R_2,\,
 x_3 + 2\pi n_3 R_3,\nn\\ &&
 e^{i \sum_{a=1}^3 n_a\a_{1a}} z_1,\,
 e^{i \sum_{a=1}^3 n_a\a_{2a}} z_2,\,
 e^{i \sum_{a=1}^3 n_a\a_{3a}} z_3).
\nn
\eear
Here $\a_{ij}$ is a matrix of phases.

It is possible to derive
the M(atrix)-model for such a compactification using the rules
of toroidal compactifications \cite{BFSS,Wati,GRT}.
A M(atrix)-model for a similar theory was discussed in
\cite{BanMot,WitPQ,CGK} and we will repeat the arguments here.
To compactify M(atrix)-theory on $\MT{3}$, we need to pick 3 
$U(\infty)$ matrices, $\Omega_1,\Omega_2,\Omega_3$
and require that the matrices satisfy \cite{BFSS,Wati,GRT}:
$$
\Omega_j^{-1} X_i \Omega_j = X_i + 2\pi \delta_{ij} R_i,\qquad
\Omega_j^{-1} Z_a \Omega_j = e^{i\a_{ja}} Z_a.
$$
The solution is to think about $\Omega_j$ and the $X_i$'s as operators
on the Hilbert space of functions on a dual $\MHT{3}$ of radii
$\frac{1}{2\pi R_j}$.
Let $0\le \sigma_j\le \frac{1}{R_j}$ (with $j=1,2,3$)
be the 3 periodic coordinates.
$\Omega_j$ is taken to be diagonal so that
$$
\langle \sigma_1\sigma_2\sigma_3 |\Omega_j 
|\sigma_1'\sigma_2'\sigma_3'\rangle = e^{2\pi i R_j\sigma_j}
\delta(\sigma_1-\sigma_1')
\delta(\sigma_2-\sigma_2')
\delta(\sigma_3-\sigma_3')
$$
and $X_j=i\px{\sigma_j} -A_j(\vsig)$ where $A_j$ is a gauge-field
on the dual $\MHT{3}$.
The twist requires us to find operators $Z_a$ such that:
$$
\Omega_j^{-1} Z_a\Omega_j = e^{i\a_{ja}} Z_a.
$$
The solution is an operator $Z_a$ with matrix elements of the form:
$$
\langle \sigma_1\sigma_2\sigma_3 |Z_a
|\sigma_1'\sigma_2'\sigma_3'\rangle = 
\delta(\a_{1a}+\sigma_1-\sigma_1')
\delta(\a_{2a}+\sigma_2-\sigma_2')
\delta(\a_{3a}+\sigma_3-\sigma_3')\Phi_a(\vsig),
$$
where $\Phi_a$ are arbitrary local fields on $\MHT{3}$.
If we did not have the twist, the Lagrangian would be 3+1D SYM
with $\SUSY{8}$ supersymmetry, as in \cite{BFSS,Wati,GRT}.
The effect of the twist is to make the fields $\Phi_a(\vsig,t)$
nonlocal.
Instead of transforming in the adjoint of the local $U(N)$
gauge group at the point $(\vsig,t)$, they transform in 
the representation $(\rep{N},\rep{\overline{N}})$ of
$U(N)_{(\vsig,t)}\times U(N)_{(\vsig+\vec{L}_a,t)}$
(where $U(N)_{(\vsig,t)}$ is the restriction of the guage group
to the point $(\vsig,t)$ and 
$$
\vec{L}_a\equiv 
\left(\frac{\a_{1a}}{2\pi R_1},
\frac{\a_{2a}}{2\pi R_2},
\frac{\a_{3a}}{2\pi R_3}\right).
$$
For example, the covariant derivative is defined as:
\be\label{DiPh}
D_i \Phi_a \equiv
\px{i}\Phi_a(\vsig) - i A_i(\vsig)\Phi_a(\vsig) 
   +i \Phi_a(\vsig) A_i(\vsig + \vec{L}_a).
\ee
Such theories where discussed in \cite{BG}, where they were obtained
by studying T-duality in gauge-theories on Noncommutative spaces,
and in \cite{DGG} from pinned-branes.
They will be referred to as ``{\bf dipole-theories}.''

Let us check that the M2-branes cannot separate, for generic twists,
$\vec{L}_a$.
To see this note that the $U(1)\subset U(N)$ does not decouple
in the dipole-theories.
In the presence of an instanton, and for generic twists,
the equation $D_i\Phi_a=0$,
with $D_i\Phi_a$ defined in (\ref{DiPh}) has no 
nonzero solution. Note that:
$$
\com{D_i}{D_j}\Phi_a = 
  -i F_{ij}(\vsig)\Phi_a(\vsig) +i \Phi_a(\vsig)F_{ij}(\vsig+\vec{L}_a),
$$
and in general $\det F_{ij}(\vsig)\neq \det F_{ij}(\vsig+\vec{L}_a)$.
Separating the M2-branes requires $\Phi_a$ to be nonzero.

In the M(atrix)-model, we should keep the quadratic terms in $\Phi_a$
and the fermions. We can neglect the quartic terms because, as we argued
above, the fluctuations in $\Phi_a$ are much smaller than
$M_p^{-1}$. The quadratic terms can be integrated to give 
factors of $\det D^i(\vec{L}_a) D_i(\vec{L}_a)$ (as a function
of the point in the instanton moduli space).
The fermion contribution does not cancel this because, generically,
supersymmetry is broken by the twists.

\section{Partition functions and Operators}
\label{maps}
We now discuss in more detail the ``dictionary'' that translates
questions about the field theories to questions
about their M(atrix)-models. For simplicity we will restrict
ourselves to the $Spin(8)$ theory.

In section (\ref{models}) we suggested that the $Spin(8)$ $SU(k)$
theory compactified on $\MT{3}$ is related to the large $N$ limit
of the integral over $\cM_{N,k}$,
but following the discussion in section (\ref{flat}),
we need to mod out the overall translation modes along 
$\MT{3}$.
We therefore define $\hM_{N,k}$ -- the ``reduced'' moduli
space of $k$ $SU(N)$ instantons on $\MT{3}\times\BR$.
By ``reduced'' moduli space we mean the following.
Let $\cM_{N,k}$ be the moduli space of $k$ instantons
of $SU(N)$. The $\MT{3}\times\BR$,
considered as an abelian group, acts on $\cM_{N,k}$ by translations.
The reduced space is the space of orbits of this $\MT{3}\times\BR$,
which we will denote by $\hM_{N,k}$.
By an ``integral'' over the moduli space, we mean the dimensional
reduction to 0D of the 0+1D Quantum-Mechanics with $8$ supersymmetries
over the moduli space of instantons.

The relation between the compactified $Spin(8)$ theory and
the integral over the instanton moduli space should be interpreted
as follows. For every operator in the $Spin(8)$ theory
there should exist an appropriate function of the moduli space
such that insertions of the $Spin(8)$
operators into the partition function correspond to insertion
of the functions into the integral.

There is no obvious reason to expect that the partition
function of the $Spin(8)$ theory itself should be equal to the 
partition function of its M(atrix)-model.
Thus, we expect to have a fixed numerical coefficient $C_{N,k}$
that is the ratio of the partition function of the $Spin(8)$ theory
and the integral over the $\hM_{N,k}$. Once this $C_{N,k}$ is 
known we can begin to map operators on the field theory side
to insertions in the M(atrix)-model integral. We will
make some remarks about $C_{N,k}$ in subsection (\ref{pf}),
but let us start with the operators.

\subsection{Momentum and the $\BZ_N$ symmetry}
Suppose we wish to calculate an expectation value:
$$
\ev{\cO_1(p_1)\cO_2(p_2)\cdots\cO_n(p_n)}
$$
in the $Spin(8)$ theory. Here $\cO_i$ are operators, for example
components of the energy-momentum tensor or of the $Spin(8)$ current
and $p_i$ are discrete momenta along $\MT{3}$.
The momenta  belong to a lattice that determines the dual $\MHT{3}$.

We conjecture that the expectation value above can be calculated
from the matrix model as:
\be\label{mmcorr}
\lim_{N\rightarrow\infty} C_{N,k}\int_{\hM_{N,k}}
 \hO_1(p_1,\xi) \cdots \hO_n(p_n,\xi) \lbrack D\xi\rbrack,
\ee
where $\hO(p,\xi)$ can be determined from $\cO(p)$ 
and $\xi\in\hM_{N,k}$.
We will not give a complete prescription to
determine $\hO$ from $\cO$ but we will suggest a treatment of
the momentum label, $p$.

Recall that in the M(atrix)-model for M-theory on $\MT{3}$,
momentum becomes electric-flux \cite{GRT} on the dual
$\MHT{3}$.
Let us present $\MHT{3}$ as $\MR{3}$ divided
by a lattice $\Gamma$. The momentum $p$ corresponds to
a vector $\vec{v}(p)\in\Gamma$ and so does electric flux.
The operators that ``have'' electric flux are the Wilson
lines. 
Let $\wxi\in\cM_{N,k}$ be a particular instanton configuration
and let $\vec{x}\in\MHT{3}$ be a point and $t\in\BR$ by a time
coordinate.
Let $W(p,\vec{x},t,\wxi)$ be the Wilson line that corresponds to
a path that is a straight line from $\vec{x}$ to $\vec{x}+\vec{v}(p)$
in $\MHT{3}$. $W$ is defined up to a gauge transformation.

We conjecture that $\hO(p,\wxi)$ has the form:
\be\label{defhO}
\int d^3 x\, dt\, \tr{\cwO(\vec{x},t,\wxi) W(p,\vec{x},t,\wxi)},
\ee
where $\cwO$ is some local operator  that is independent of 
$p$.

Note that because we integrate over $\MHT{3}\times\BR$,
$\hO(p,\wxi)$ depends only on the equivalence class of $\wxi$
under translations, i.e. on $\xi$.

Is momentum conserved according to this definition?
The answer is yes, but only in the large $N$ limit!

To see this note that as defined, $\hO$ has a $\BZ_N$
ambiguity. This is because an instanton configuration defines
an $SU(N)/\BZ_N$ gauge configuration (since gauge fields
are in the adjoint representation) but the Wilson-line traces
are defined in the fundamental representation. Therefore,
the Wilson lines have an ambiguity that corresponds to making
a $\BZ_N$ gauge transformation in the center of $SU(N)$ as we
go along any of the  1-cycles of $\MHT{3}$.
The  integral in (\ref{mmcorr}) will be independent of 
this $\BZ_N$ ambiguity only if the sum of the overall
momenta of all the operators is an $N$-multiple of a $\Gamma$-lattice
vector. In the large $N$ limit this could only be zero.

To summarize, the definition (\ref{defhO}) satisfies:
\begin{itemize}
\item
The nontrivial classes of 1-cycles on $\MHT{3}$ are naturally
labeled by the momentum, $p$, on the original $\MT{3}$.
Therefore, mapping an operator that carries $p$ units of momentum
to an operator that contains a Wilson line along an appropriate
1-cycle is natural.
\item
By introducing Wilson lines in the fundamental representation we have 
introduced a $\BZ_N^3$ symmetry. The $\BZ_N^3$ quantum numbers of
$\hO(p,\xi)$ should equal $p\pmod N$.
\item
Momentum is conserved up to multiples of $N$ because
of the $\BZ_N^3$ ambiguity in determining the Wilson line traces.
\item
By integrating over the whole $\MHT{3}\times\BR$  the
operator becomes independent of the center-of-mass position
of the instanton configuration, as is required by our prescription
of integrating over the reduced moduli space.
\end{itemize}

If the above conjecture is correct then it is enough to determine
the local variables, $\cwO$, by determining the mapping of operators
with zero momentum. We will therefore
turn now to discuss the zero modes of the energy momentum tensor
and R-symmetry currents.

\subsection{Energy momentum tensor}
The M(atrix)-variable, $\hO$, corresponding to an
energy momentum tensor insertion, $T_{\u\v}(\vec{0})$,
with zero momentum
can be deduced by studying the variation
in the volume form on $\hM_{N,k}$ as a result
of an infinitesimal change in the metric on 
$\MHT{3}$. We will not do this here, but instead we will
sketch the procedure for a more complicated operator insertion.

The $Spin(8)$ theory has a Noether current operator, $J^a_\u$
($\u=0,1,2$ and $a=1\dots 28$, the dimension of $Spin(8)$)
corresponding to the global $Spin(8)$ symmetry.
The construction in section (\ref{models}) makes only an
$SU(4)=Spin(6)\subset Spin(8)$ manifest.
We will now map the integral, $\int_{\MT{3}} J^a_\u(x) d^3 x$,
($a=1\dots 15$), of the current corresponding to that $Spin(6)$.

For simplicity, let us consider a rectangular 
$\MT{3}=\MS{1}\times\MS{1}\times\MS{1}$. Let the radii of the $\MS{1}$'s
be $R_1,R_2,R_3$ and let us map the component of the current along
the first $\MS{1}$.
Suppose we add an infinitesimal term,
$\epsilon J^a_\u$, to the Lagrangian of the 3D $Spin(8)$
theory. In M-theory, this can be realized as 
a geometrical twist in the boundary conditions for $\MS{1}$.
Thus, when we go once around $\MS{1}$, we also perform
a $Spin(6)$ rotation in the transverse $\MR{6}$ (keeping
$\MR{1,1}$ that contains the light-like direction intact).

We have described the M(atrix)-models for such twisted compactifications
in subsection (\ref{subtwist}).
We have argued that the modification to the integral over the moduli
space of instantons, because of the twists, is given
by integrating the quadratic term in the adjoint bosons and fermions.
The bosons give ${\mbox{det}}^{-1} (D^i(\vec{L}) D_i(\vec{L}))$,
where $D^i(\vec{L})$ is the $\vec{L}$-dependent covariant
derivative, defined in (\ref{DiPh})
and the fermions give
${\mbox{det}} (\Dsl^i(\frac{\vec{L}}{2}) \Dsl_i(\frac{\vec{L}}{2}))$.
For $\vec{L}=0$ the fermions have zero-modes that
become the superpartners of coordinates on the moduli space
of instantons. For generic $\vec{L}\neq 0$, there are no zero modes
as we have seen in (\ref{subtwist}).
Expanding the expressions for small $\vec{L}$ should in principle
determine the operator insertion.

\subsection{The partition function of the $Spin(8)$ theory}\label{pf}
We will now return to the point that was left open at the beginning
of the section, namely the coefficient $C_{N,k}$ -- the ratio
between the $Spin(8)$ partition function and its M(atrix)-model's
partition function.
The integral over the moduli space of instantons, with the fermions,
becomes the integral of the Euler density:
$$
J_{N,k}\equiv \int_{\cM_{N,k}} e(\cM_{N,k}),
$$
We wish to compare it to the partition function of the $Spin(8)$
theory and find the numerical relation between the two quantities.

The partition function of the $Spin(8)$ theory compactified
on $\MT{3}$ is given by:
$$
I_k = \sum_{d|k} \frac{1}{d}.
$$
where the sum is over all divisors $d$ of $k$.

This can be argued as follows.
Recall that
the partition function for the D-instanton action (10D $SU(k)$
Super-Yang-Mills
dimensionally reduced to 0D) is given by $\sum_{d|k} \frac{1}{d^2}$.
For $k=2$ this was calculated in \cite{Yi,SetSte} and it was then
derived from the conjectured type-IIB $R^4$-coupling 
\cite{GG} (these couplings were proven in \cite{GreSet},
and see also \cite{PioR}.)
A similar calculation can be performed for the effective $R^4$
coupling in M-theory compactified on $\MT{3}$. The result is \cite{PioKir}:
$$
\sum_{k=0}^\infty q^k I_k  = \log \prod_{n=1}^\infty \frac{1}{1-q^n},
$$
from which the expression for $I_k$ follows.

We will not calculate the integral $J_{N,k}$ for
$\MT{3}\times\BR$ in these notes but we will note
the following special limit.
If we replaced $\MT{3}$ with $\MR{3}$,
we could then borrow the known
results for the integral over the instanton moduli space
on $\MR{4}$. In \cite{DHK} the following result is given:
$$
\lim_{N\rightarrow\infty} J_{N,k}(\MR{4})
 = 2^{3-2k}\pi^{6k-13/2}\sqrt{N} k^{3/2}\sum_{d|k} \frac{1}{d^2}.
$$
In order to compare it to the partition function
of the $Spin(8)$ theory we need to decompactify
the $\MHT{3}$ that appears on the M(atrix)-theory side.
This can be done along the following steps.
The $Spin(8)$ theory compactified on $\MS{1}\times\MR{2}$
with the $\MS{1}$ of radius $r$ can be described at low-energies
by 2D SYM with $\SUSY{(8,8)}$ supersymmetry and coupling constant
that is proportional to $r^{-1/2}$. 
Replacing $\MHT{3}$ by $\MHT{2}\times \BR$ on the M(atrix)-theory
side corresponds to replacing the $Spin(8)$ theory with 2D SYM
compactified on $\MT{2}$ and replacing $\MHT{3}$ by $\MR{3}$
corresponds to replacing the partition function
of the $Spin(8)$-theory on $\MT{3}$ with the D-instanton integral
of 10D SYM reduced to 0D. This was calculated in
\cite{GGii} to give $\sum_{d|k}\frac{1}{d^2}$ (see also
\cite{Yi,SetSte}). So the proportionality
coefficient in this case would be
 $C_{N,k} 
  = 2^{2k-3}\pi^{\frac{13}{2}-6k}N^{-\frac{1}{2}} k^{-\frac{3}{2}}$.

\section{Extension to noncommutative spaces}\label{noncom}
$\SUSY{4}$ SYM on a noncommutative $\MR{3,1}$
(NCSYM) can be realized in string-theory  by turning
on a strong NSNS 2-form $B$-field on the brane \cite{CDS,DH,SWNCG}.
The NCSYM theories are labeled by an anti-symmetric 
contravariant 2-tensor (bivector) $\th^{ij}$.
The standard M(atrix)-models for $\SUSY{4}$ SYM can be extended
to describe NCSYM, but not all the components of $\th^{ij}$ can
easily be turned on, as we will explain below.
On the other hand, all 6 components of $\th^{ij}$ on $\MT{4}$
can easily be turned on in the new M(atrix)-model.

\subsection{Noncommutativity in the standard M(atrix)-models}
In \cite{GanSet}, the M(atrix)-models for
$\SUSY{4}$ SYM were derived from the M(atrix)-models of the $(2,0)$
theory \cite{ABKSS,WitQHB}.
These M(atrix)-models where a special limit of the moduli space
of instantons on $\MT{2}\times\MR{2}$. In this limit the $\MT{2}$
becomes small and the moduli space can be described as the moduli
space of holomorphic curves in $\MT{2}\times\MR{2}$.

In \cite{ABS}, a ``noncommutative''
extension of the 5+1D $(2,0)$-theory was suggested.
It was parameterized by a 2-form $C_{ij}$ that is anti-self-dual
in 4 space-like directions.
Its M(atrix)-model was described as Quantum-Mechanics on a certain
deformation of $\cM_{N,k}(\MR{4})$. This target space was  later
identified as the extension of $\cM_{N,k}$
to instantons on a noncommutative $\MR{4}$ \cite{Berk,NekSch}.

Given these constructions, it is not hard to extend the results
of \cite{GanSet} to NCSYM as long as 4 out of the
6 components of $\theta^{ij}$ are set to zero.
When we discuss the ``standard'' M(atrix)-model of $\SUSY{4}$
SYM on $\MR{3,1}$, we have to pick two light-like directions
whose coordinates are
denoted by $x_\pm \equiv x_0\pm x_3$ (see \cite{BanksRev,WatiRev} for
a review of M(atrix)-theory). Let the coordinates of the 
other two directions be denoted by $x_1,x_2$.
The dual light-like momenta are $p^{+} = N/R_\|$ and $p^-$ 
(that equals the Hamiltonian).
In this notation only $\theta^{1+},\theta^{2+}$ can be turned on.
The resulting M(atrix)-model is a limit of the moduli space of
instantons on a noncommutative $\MT{2}\times\MR{2}$.
The noncommutativity is given by an anti-self-dual
 2-tensor with one direction along $\MT{2}$ and the other along $\MR{2}$.

To obtain the M(atrix)-model for NCSYM one has to proceed along
the lines of \cite{GanSet} and take the limit of a small $\MT{2}$.
In this limit, the commutative moduli space reduces to 
a certain moduli-space of holomorphic curves inside $\MT{2}\times\MR{2}$.
In the noncommutative case one also obtains a moduli space
of curves, but the space $\MT{2}\times\MR{2}$ has to be deformed
in the following way \cite{GaMiSa}
(see also \cite{AsNeSc} for general properties
of instantons on noncommutative tori).
Let $w$ be a holomorphic coordinate on $\MT{2}$ and let
$z$ be a holomorphic coordinate on $\MR{2}$.
In the commutative case, the periodic identification is
$w\sim w+n+m\tau$, ($n,m\in\BZ$ and $\tau$ is the complex structure
of $\MT{2}$). In the noncommutative case, 
the identification is deformed into:
$$
(w,z)\sim 
(w + n + m\tau, z-{{2\pi i}\over{\tau_2}}(\th^{1+}+i\th^{2+})
   (n+m\overline{\tau})),
\qquad ,n,m\in\BZ.
$$
Thus, completing a cycle around $\MT{2}$ has to be accompanied by
a translation along $\MR{2}$.
It is not clear how to turn on other components of $\th$ in the
M(atrix)-model.

\subsection{Noncommutativity in the new M(atrix)-models}
In contrast to the discussion in the previous subsection,
in the new M(atrix)-models all 6 noncommutativity parameters enter
on an equal footing.
To turn on noncommutativity we need to turn on a strong
NSNS 2-form $B$-field along $\MT{4}$ and take
a scaling limit in which the size of the $\MT{4}$
shrinks to zero but the $B$-field flux remains finite \cite{SWNCG}.

Alternatively, we can start with the definition of \cite{DH,CDS}
of the NCSYM limit on tori. 
We therefore start with
type-IIB string theory on $\MT{4}$ with radii $R_1,\dots,R_4$
such that (in the notation of subsection (\ref{mmsym}))
$M_s R_i\rightarrow 0$. We turn on an NSNS 2-form $B$-field 
with a finite flux along $\MT{4}$.
We then need to look for an instanton 
that has the charge of $N$ D$(-1)$-branes.
According to the arguments of \cite{DH}, in this limit,
the low-energy description is $\SUSY{4}$ SYM on the T-dual
$\MT{4}$ with noncommutativity that is set by the $B$-field fluxes.

To find the M(atrix)-model, we follow the steps described in subsection
(\ref{mmsym}) and obtain  LST on $\MT{2}\times\MHT{3}$.
 Looking at the formulas (\ref{ATtwo})-(\ref{LiRad}) we see
that the size of $\MHT{3}$ is finite, in little-string units,
but the size of $\MT{2}$ shrinks to zero.
The 6 components of the NSNS 2-form B-field become the components
of an external NSNS 2-form $B$-field for the LST with one 
index along $\MT{2}$ and another along $\MHT{3}$.
Since the $\MT{2}$ is small we need to perform another T-duality
to make it big. The $B$-field fluxes become components of
the metric with one index along $\MT{2}$ and another along
$\MHT{3}$. More precisely, they become Dehn twists in the 1-cycles
of the $\MHT{3}$ as we go along 1-cycles of $\MT{2}$.
At the end of the duality transformations,
the number of D$(-1)$-branes
becomes the instanton number on $\MHT{3}\times\BR$, as is clear
from the case when no fluxes are turned on.

We therefore end up with a similar $\sigma$-model to the one
we had in subsection (\ref{mmsym}) except that we do not
mod out by the translations along $\MHT{3}$ and we need to
introduce boundary conditions along $\MT{2}$ that correspond
to translations along $\MHT{3}$ as we go around a 1-cycle
of $\MT{2}$.
These boundary conditions are described by 6 parameters,
3 for each 1-cycle of $\MT{2}$. These parameters are proportional
to the 6 noncommutativity parameters along the original $\MT{4}$.
The need to retain the translations along $\MHT{3}$  is probably
related to the non-decoupling of the center of $U(N)$ in
NCSYM (see \cite{SWNCG}).

\subsection{S-duality} 
One of the pleasing features of the ``standard'' M(atrix)-model
of $\SUSY{4}$ SYM is that S-duality is manifest \cite{GanSet}.
S-duality can be extended to NCSYM and takes $\th^{ij}$ to
its dual tensor ${\epsilon^{ij}}_{kl}\th^{kl}$ \cite{GMMS,GGS,SSTii}.
On a Euclidean space the S-dual theory is well-defined,
but on $\MR{3,1}$ there are complications related to space-time
noncommutativity \cite{GMMS,GGS,SSTii}
and more degrees of freedom are required to
make the theory consistent \cite{GMMS,SSTii} (and see also
the related discussions in \cite{SSTi}-\cite{AhGoMe}).

In the new M(atrix)-models S-duality is also manifest.
It is just the $SL(2,\BZ)$ duality of the base $\MT{2}$
on which the $\sigma$-model is defined.
The S-duality that acts on the coupling as
$\tau\rightarrow -1/\tau$  exchanges the boundary conditions
along the short cycle of $\MT{2}$ with the boundary conditions
along the long cycle of $\MT{2}$.
Following the steps of (\ref{mmsym}),
this can be seen to agree with taking $\th^{ij}$ to its dual tensor.

\section{Summary}
We have proposed that $\SUSY{4}$ $SU(k)$ SYM compactified
on $\MT{4}$ with coupling constant
$\tau=\frac{4\pi i}{g^2} + \frac{\theta}{2\pi}$ has a manifestly S-dual
M(atrix)-model given by the large $N$ limit of a $\sigma$-model
compactified on $\MT{2}$ with complex structure $\tau$.
The target space of the $\sigma$-model is the moduli space $\cM_{N,k}$
of $k$ $SU(N)$ instantons on  $\MHT{3}\times\BR$.
The parameters (moduli) of the $\sigma$-model are determined
from the shape of $\MT{4}$ (see section (\ref{models})).
The M(atrix)-model for $\SUSY{4}$ $U(k)$ SYM on a noncommutative
$\MT{4}$ corresponds to a similar $\sigma$-model but
with modified boundary conditions along $\MT{2}$.
The modification is that as we go around a 1-cycle of $\MT{2}$
the instanton configuration is shifted by a translation
along $\MHT{3}$.

We have also proposed that the M(atrix)-model
of the $Spin(8)$ theory (the CFT
associated with $k$ M2-branes)  compactified on $\MT{3}$
is an integral over $\cM_{N,k}$.

We conjectured that operator insertions in the field-theory
that carry a specific momentum along $\MT{3}$ correspond
to Wilson line insertions in the M(atrix)-model where the Wilson
line is calculated along a cycle on the dual $\MHT{3}$ that
is related to the momentum.
Momentum conservation is achieved in the large $N$ limit
due to the $\BZ_N$ ambiguity of the Wilson line.

These results were derived directly from M(atrix)-theory but
it might be interesting to derive them also from the AdS/CFT
correspondence, along the lines of \cite{DHKMVi,DHKMVii,DHK}.
For example, the $\SUSY{4}$ SYM appears on a Euclidean D3-brane
wrapping $\MT{4}$ in $\AdS{3}\times \MS{3}\times\MT{4}$
and the latter is believed to 
be dual to the large $N$ limit of a 2D CFT \cite{M,GKS,KutSei}.
It would be interesting to study the instantons in that picture.

\section*{Acknowledgments}
I am indebted to Nikita Nekrasov for helpful discussions
about Euler numbers of instanton moduli spaces.
This research is supported by NSF grant number PHY-9802498.

\def\np#1#2#3{{\it Nucl.\ Phys.} {\bf B#1} (#2) #3}
\def\pl#1#2#3{{\it Phys.\ Lett.} {\bf B#1} (#2) #3}
\def\physrev#1#2#3{{\it Phys.\ Rev.\ Lett.} {\bf #1} (#2) #3}
\def\pr#1#2#3{{\it Phys.\ Rev.} {\bf D#1} (#2) #3}
\def\ap#1#2#3{{\it Ann.\ Phys.} {\bf #1} (#2) #3}
\def\ppt#1#2#3{{\it Phys.\ Rep.} {\bf #1} (#2) #3}
\def\rmp#1#2#3{{\it Rev.\ Mod.\ Phys.} {\bf #1} (#2) #3}
\def\cmp#1#2#3{{\it Comm.\ Math.\ Phys.} {\bf #1} (#2) #3}
\def\mpla#1#2#3{{\it Mod.\ Phys.\ Lett.} {\bf #1} (#2) #3}
\def\jhep#1#2#3{{\it JHEP.} {\bf #1} (#2) #3}
\def\atmp#1#2#3{{\it Adv.\ Theor.\ Math.\ Phys.} {\bf #1} (#2) #3}
\def\jgp#1#2#3{{\it J.\ Geom.\ Phys.} {\bf #1} (#2) #3}
\def\cqg#1#2#3{{\it Class.\ Quant.\ Grav.} {\bf #1} (#2) #3}
\def\hepth#1{{[hep-th/{#1}]}}


\end{document}